\documentstyle[procl]{article}

\input{psfig.sty}

\def\Journal#1#2#3#4{(#1) {#2} {\bf #3}, #4}

\newcommand{\be}{\begin{equation}}
\newcommand{\ee}{\end{equation}}
\newcommand{\ba}{\begin{eqnarray}}
\newcommand{\ea}{\end{eqnarray}}

\def\AAp{\em Astron. Astrophys.}
\def\AJ{\em Astron.~J.}
\def\ApJ{\em Astrophys.~J.}

\def\AaAp{\em Astron. Astrophys.}

\def\MNRAS{\em Mon. Not. R.~Astron. Soc.}
\def\Nat{\em Nature\/}

\begin{document}

\title{Statistics of Turbulence from Spectral-Line Data Cubes}

\author{Alex Lazarian}

\address{ Department of Astronomy, University of Wisconsin,
Madison, USA; email: lazarian@astro.wisc.edu}

\maketitle

\abstract{Emission in spectral lines can provide unique information
on interstellar turbulence. Doppler shifts due to supersonic 
motions contain information on turbulent velocity field which is
otherwise difficult to measure. However, the problem of separation
of velocity and density fluctuations is far from being trivial. Using atomic
hydrogen (HI) as a test case, I review
techniques applicable to emission line studies with the emphasis on those
that can provide information on the underlying {\em power spectra} of
velocity and density. I show 
that recently developed mathematical machinery is promising
for the purpose. Its application to HI
 shows that in cold neutral hydrogen the velocity fluctuations dominate
the small scale structures observed in spectral-line 
data cubes and this result is
very important for the interpretation of observational data, including
the identification of clouds. Velocity fluctuations are shown to 
dominate the formation of small scale structures that can be erroneously
identified as diffuse clouds.
One may argue that the HI
data is consistent with the
Goldreich-Shridhar picture of magnetohydrodynamic
turbulence, but the cascade from the scales of several kpc that this
interpretation involves
does not fit well in the current paradigm of energy injection. 
The
issue whether magnetic field does make the turbulence anisotropic
is still open, but if this is the case, I show that studies of emission
lines can provide a reliable way of determining magnetic field direction.
I discuss various techniques for studying 
interstellar turbulence
using emission lines, e.g. spectral correlation functions, genus
statistics and principal component analysis.   
}

   \section{Introduction}

The interstellar medium  is turbulent and the turbulence is crucial for
understanding of various interstellar processes. Interstellar turbulence
occurs in magnetized fluid and magnetic field establishes a connection
between ISM phases (McKee \& Ostriker 1977) thus making the turbulent
cascade much more complex and coupling together cosmic rays and gas. 
Theoretical understanding of such a multiphase media with the
injection of energy at different scales (Scalo 1987) is extremely
challenging.

In terms of the topic of the present meeting, turbulence is
important both for accelerating cosmic rays and for their diffusion.
Indeed, whatever mechanism of cosmic ray acceleration we consider,
its understanding requires proper accounting for scattering
of cosmic rays by turbulent magnetic field. The same is 
true for the propagation of cosmic rays. For instance, if it were not 
for magnetic field
lines wandering, the diffusion of cosmic rays perpendicular to the
magnetic
field direction would be suppressed (see Jokipii 1999). Moreover,
it is becoming clear that particle streaming along magnetic field lines
is also substantially influenced by magnetic turbulence. 

In view of a broader picture, turbulence is widely believed to
be an important element of molecular cloud dynamics 
and star formation process, although various 
authors disagree on the degree of its importance (see discussion
in Vazques-Semadeni
\& Passot 1999). Undoubtedly turbulence is essential for heat 
transfer in the interstellar medium. It has been recently
suggested that turbulence
is also 
a key element to understanding various chemical reactions (Gredel 1999)
and of the fundamental problem of MHD, namely, to the problem of
fast magnetic reconnection (Lazarian \& Vishniac
1999). This very limited and incomplete list of processes for which
turbulence is essential explains the motivation behind the attempts
to study interstellar turbulence.

Unfortunately interstellar turbulence remains a mystery in spite of all the
attempts to study it. Substantial progress in numerical research (see
Ostriker 1999, Vazquez-Semadeni \& Passot 1999) is not adequate to reproduce
the flows comparable in complexity and in Reynolds numbers, and the
situation will not change 
in any foreseeable future. Thus only direct observational 
studies of interstellar
turbulence may provide us with the crucial
information on this phenomenon. Approaching the problem one would like
to know at least the statistics of density, velocity and magnetic field.
In this review I briefly discuss what information emission lines can
supply us with. I would like to quote Alyssa Goodman, who believes
that present day technology made spectral-line
mapping of large portions of interstellar media 
``a booming cottage industry''. Attempts to use this wealth of observational
data via visual inspection become fruitless and this calls for the 
introduction of more sophisticated techniques.

Statistical description is a nearly indispensable strategy when
dealing with turbulence and a big advantage of statistical techniques
is that they extract underlying regularities of the flow and reject
incidental details. 
Attempts to study interstellar turbulence with statistical tools
date as far back as the 1950s
(see Horner 1951, Kampe de Feriet 1955, Munch 1958, 
Wilson et al. 1959) and various directions
of research achieved various degree of success (see reviews by
Kaplan \& Pickelner 1970, Dickman 1985, Lazarian 1992, Armstrong, Rickett
\& Spangler 1995). 
Studies of turbulence statistics of ionized media were successful
(see Spangler \& Gwinn 1990, Narayan 1992) and provided the information of
the statistics of plasma density\footnote{Incidentally the found
spectrum was close to a Kolmogorov one.} at scales $10^{8}$-$10^{15}$~cm. 
This research profited
a lot from clear understanding of processes of scintillations and scattering
achieved by theorists (see Goodman \& Narayan 1985, Narayan \& Goodman 
1989). At the same time 
the intrinsic limitations of the scincillations technique
are due to the limited number of sampling directions and difficulties
of getting velocity information.

Deficiencies in the theoretical description have been, to our mind, the
major impediments to studies of turbulence using emission lines. 
For instance, important statistical studies of molecular clouds 
(Dickman 1985, Dickman \& Kleiner 1985, Miesch \& Bally 1994) have not
achieved the success parallel to that in scintillation studies. 

Potentially, studies of interstellar turbulence via emission lines
can provide statistics of turbulence in various 
interstellar phases, including neutral gas. More importantly, velocity
information allows one to distinguish between
static structures and  dynamical turbulence.

The difficulty of studying Doppler broadened lines stems from the fact
that
one has to account for both velocity and density fluctuations. Indeed,
at any given velocity the fluctuation of emissivity may arise both
from the actual blobs of gas moving at this velocity and from parcels
of gas with different spatial positions but accidentally having the same
component of velocity along the line of sight. Therefore fluctuations
of emissivity at a given velocity would be expected even if the media were
completely incompressible. 

There exist various ways of dealing with position-position-velocity 
(henceforth
PPV)
data cubes. One of them is to identify clumps and to describe their
statistics (see Stutzki \& Gusten 1990, Williams, de Geus \& Blitz 1994).
Another is use a more traditional set of hydrodynamic tools like
power spectra, structure functions etc. The two statistics are interrelated
(see Stutzki et al. 1998), but in general the relation between various
tools is non-trivial. It seems that for answering various
questions different statistical tools are more suitable. Therefore
it is very encouraging that a number of techniques, including Principal
Component Analysis (see Heyer \& Schloerb 1997) and Spectral Correlation
Functions (Goodman 1999, Rosolowsky et al. 1999) have been 
recently introduced to the field. 

In what follows I depart from a traditional statistical hydrodynamics and
describe how the 3D velocity and density power spectra can be extracted from
position-position-velocity (PPV) data cubes. This choice reflects my subjective
preferences and partially motivated by the fact that this approach relates the
long-studied 3D density and velocity statistics (e.g. power spectra)
with the observational data. Even with this limitation  the scope the subject
is too broad and I shall mostly talk about atomic hydrogen (HI) studies,
that can be viewed as a test case for the technique. I discuss advantages
of using HI as a test case in section~2, the problem of space-velocity
mapping in section~3 and spectra in velocity slices in section~4.
The interpretation of 21~cm Galactic and SMC
data is given in section~5. Possible anisotropies of statistics 
stemming from magnetic
field are dealt with in section~6, where a new technique for statistical 
studies of magnetic field is suggested. I consider formation of
emissivity enhancements that can be identified as 
 filaments and clouds in section~7 and 
discuss the generalization of the technique in section~8. Being aware of
the limitations of the traditional hydrodynamic description of turbulence,
we  describe alternative approaches, i.e.
 2D Genus statistics, Spectral Correlation Functions and
Bispectrum in section~9. A short discussion of the results is given
in section~10.

\section{HI as a Test Case} 

Atomic hydrogen is an important component of the interstellar media 
and many
efforts have been devoted to its studies (see Burton 1992). 
In terms
of turbulence studies it has a number of advantages. For one thing, when
dealing with HI one may in most cases disregard self-absorption.
There are two major reasons for that: self-absorption
is small (Braun 1997, Higgs 1999) and as shown in Lazarian (1995, henceforth
L95), small localized absorption features typical to HI only
marginally influence the statistics on the scales larger than their size.
For another thing, the pervasive distribution of neutral hydrogen
presents a sharp contrast to the localized distribution of
molecular species, and this alleviates problems related to averaging.
Moreover, atomic hydrogen emissivity is proportional to the
first power of atomic density and this simplifies the analysis.

HI has a substantial filling factor ($\sim 20\%$ or larger) in the Galactic
disc and therefore its motions should reflect large scale galactic
supersonic turbulence. At the same time, its statistics 
may have connection with the statistics of molecular clouds. An
additional advantage of HI is that it can be studied not
only within our Galaxy but for the nearby galaxies as well.

Another motivation for studies of HI statistics stems from the recent attempt
to describe the structures in the Galactic hydrogen in order
to estimate the fluctuations of microwave polarization arising from
interstellar dust. This contribution is extremely important in view of
present-day efforts in the Cosmic Microwave Background (CMB)
 research (see Prunet \& Lazarian 1999,
Draine \& Lazarian 1999). Some of the studies, for instance 
one by Sethi, Prunet and Bouchet (1998),
attempts to relate the statistics of density observed in the velocity
space and the statistics of polarization fluctuations. If such a
relation were possible, it would greatly alleviate the efforts to study
polarization of cosmological origin. As an intermediate step in this work,
however, one should relate the statistics in emissivity in the PPV
space and density of HI in real space.
 
The timing for developing statistical tools for HI studies
is also influenced by the fact that new large data cubes, e.g.
the Canadian Galactic Survey data (see Higgs 1999), 
should become available soon.

\section{Basic equations}

\subsection{Space-Velocity Mapping}

{\em Problem}\\
The notion that the velocity fluctuation can influence emissivity
within PPV data cubes is not new.
Since the early-seventies 
Butler Burton on numerous occasions claimed the importance of
velocity fluctuations for the interpretation of 21~cm data (Burton 1970,
1971). The
ambiguities of inferring cloud properties from CO emission lines
were discussed by Adler \& Roberts (1992). Using N-cloud
simulations of spiral disks they showed that many spurious
effects appear because of velocity blending along line of sight.
Recently a number of researchers
doing numerics (Pichardo et al. 1999, Vazques-Semadeni 1999) pointed out
that pixel-to-pixel correlation between the channel maps and 
the velocity slices of PPV data cubes tends to be larger with the
velocity rather than the  density field.

To describe power spectra of velocity and density fields, i.e. to
express the interstellar statistics using the language that was so
successful in hygrodynamics (Monin \& Yaglom 1972), one needs
to disentangle velocity and denisty contributions to 21~cm emissivity
fluctuations.

{\em Approach}\\
A quantitative 
treatment of the effects of space-velocity mapping is given in Lazarian
\& Pogosyan (1999, henthforth LP99). There it is assumed that the 
velocity of a gas
element can be presented as a sum of the regular part ${\bf v}^{reg}$
which can arise for instance from 
Galactic rotation, and a random, turbulent,
part ${\bf u}$, so that
${\bf v}^{obs} = {\bf v}^{reg} + {\bf u}$. 
The mapping from real space to PPV coordinates corresponds to
a transformation
\begin{eqnarray}
{\bf X_s} &=& {\bf X} \nonumber  \\ 
z_s &=& A \left[ f^{-1} z - {\bf u}({\bf x}) \cdot {\bf \hat z} \right]~~~, 
\label{eq:mapnoz}
\end{eqnarray}
 where henceforth we use large letters to denote vectors in the
Position-Position plane (i.e. $xy$-plane)
and use $z_s$ for  the velocity coordinate. The parameter 
$A$ is just a conversion factor which specifies the units
of $z_s$ coordinate and it is intuitively clear that this factor should
not enter any final expressions for turbulence statistics. On the contrary,
the shear parameter $f=\left(\delta v_z^{reg} / \delta z \right)^{-1}$
is an important characteristic of the mapping and one expects it
to influence our final results. For Galactic disc mapping it is convenient to choose
$A=f$, while studies of isolated clouds correspond to a zero shear,
i.e. $ f^{-1} \to 0$. As most work on HI has been done so far on Galactic
disc HI, to simplify our presentation we use the former choice. With this
definition of space-velocity mapping LP99 obtain the power spectrum
$P_s$ in the PPV space:
\begin{eqnarray}
\langle \rho_s({\bf k}) \rho_s^{*}({\bf k^{\prime}}) \rangle &=& P_s({\bf k}) 
\delta({\bf k - k^{\prime}})\nonumber \\
P_s({\bf k}) &=& e^{-f^2
k_z^2 v_T^2}
\int d^3 {\bf r} \, e^{i {\bf k}  \cdot {\bf r}} 
\Xi({\bf k}, {\bf r}),
~~~~~~~ {\bf r}={\bf x} - {\bf x \prime}~~,
\label{eq:roman}
\end{eqnarray}
where the kernel is
\be
\Xi({\bf k}, {\bf r})=\langle e^{i f 
k_z (u_z({\bf x})- u_z({\bf x \prime}))}
\rho ({\bf x}) \rho ({\bf x \prime}) \rangle~~~.
\label{kernelXi}
\ee
In derivation of this expression it is explicitly assumed that the
turbulence is statistically homogeneous in the real space coordinates
and the average denoted by angular brackets $\langle....\rangle$
depends only on the vector separation between points. The density Fourier
modes in PPV space $\rho_s({\bf k})$ are uncorrelated\footnote{A treatment
of turbulence within individual clouds is slightly different (LP99).} 
which is reflected
in $\delta$ function presence in the right-hand side of the first equation
in (\ref{eq:roman}). The factor $ e^{-f^2 k_z^2 v_T^2}$, where $v_T$
is a thermal velocity of atoms originates from averaging over thermal
distribution and it shows that only supersonic motions are readily 
available for statistical studies. Note, that
expressions similar to (\ref{eq:roman})
and (\ref{kernelXi}) were earlier obtained by Scoccimarro et al. (1999)
in the framework of studies of Large Scale Structure of the Universe
and this confirms the similarity of the problems studied in the two fields.
However, the problem of ``redshift-space'' corrections to the statistics
of galaxy distribution (Kaiser 1987)
has been addressed either in the linear regime
when perturbations are small  or when the velocity contribution to the
Fourier spectrum can be factorized by a Maxwellian factor (see Hamilton 1998).
The problem that is dealt in turbulence case is much richer as one
has to deal with non-linear density fields transformed by coherent velocities.

Note that velocity and density enter eq.~(\ref{kernelXi}) in a different
way: velocity is in the exponent and density enters as the product
$\rho ({\bf x}) \rho ({\bf x \prime})$. This provides an opportunity
to disentangle the two contributions.

\subsection{Spectrum in PPV Space}

LP99 proves that in terms of final results
for Lognormal distribution of density fluctuations
and Gaussian distribution of velocity fluctuations it is safe to
separate velocity and density in the following way
\be
\langle e^{i f \ldots} \rho ({\bf x}) \rho ({\bf x \prime}) \rangle =\langle e^{i f \ldots} \rangle \langle \rho ({\bf x}) \rho ({\bf x}+{\bf r}) \rangle~~~,
\label{sep} 
\ee
even if density and velocity are strongly correlated. 
It is interesting to check the degree of uncertainty that
 the assumption (\ref{sep}) entails using numerically generated density
and velocity fields. 

For the sake of simplicity the 
density correlation function and velocity correlation tensor
are considered to be isotropic in Galactic 
coordinates ($xyz$ space), i.e.
\be
\langle \rho ({\bf x}) \rho ({\bf x}+{\bf r}) \rangle=\xi(r)=\xi({\bf r})~~~.
\label{xifirst}
\ee 
\begin{equation}
\langle \Delta u_i \Delta u_j \rangle = \left( D_{LL}(r)-D_{NN}(r) \right) {r_i r_j \over r^2}
+D_{NN}(r) \delta_{ij}~~~,
\label{struc}
\end{equation}
where $D_{LL}$, $D_{NN}$ are longitudinal
and transverse correlation functions respectively (Monin \& Yaglom 1972), and 
$\delta_{ik}$ equals 1 for $i=k$ and zero otherwise. These assumptions
are not necessary, as the treatment can be provided for instance
for axisymmetric turbulent motions (see Oughton, Radler \& Matthaeus 1997)
as it is discussed in (L95). 

The general expression for the 3D spectrum in PPV space
is
\begin{equation}
P_s({\bf k}) = e^{-f^2 k_z^2 v_T^2/2}
\int d^3 {\bf r} \, e^{i {\bf k}  \cdot {\bf r}} \xi(r)
\exp\left[-\frac{1}{2}f^2k_z^2 D_z({\bf r})
\right],
\label{eq:main}
\end{equation}
where 
\begin{equation}
D_z({\bf r}) \equiv \langle \Delta u_i \Delta u_j \rangle \hat z_i 
\hat z_j = D_{NN}(r) + [ D_{LL}(r)-D_{NN}(r)] \cos^2\theta~,
~~~ \cos\theta \equiv {\bf \hat r \cdot \hat z}
\label{eq:Dz}
\ee
is the projection of structure tensor to the $z$-axis. 
Expression (\ref{eq:main})
is quite general and can be used to relate arbitrary velocity and density
statistics in galactic coordinates with the HI emissivity in the PPV space.

\section{Spectra in Velocity Slices}
  
Observations of Galactic HI (Green 1993) revealed two dimensional spectrum
of intensity fluctuations (see L95) and this spectrum shows power-law
behaivior. Similar power laws for Galactic data were
 obtained  by Crovisier \& Dickey (1983),
Kalberla \& Mebold (1983), Kalberla \& Stenholm (1983) and for
Small Magellanic Clouds (SMC) by Stanimirovic et al. (1999).
Thus LP99 considered power law statistics, namely, of velocity
${\cal P}_{3v}\sim k^{\nu}$ and density ${\cal P}_{3\rho}\sim k^{n}$,
where ${\cal P}$ is used to denote spectra in Galactic coordinates. Note, 
that $n<0$ and $\nu<0$ and $D_z\approx Cr^{m}$, where $m=-\nu-3$.
Power law spectra were also reported for molecular $^{12}$CO
(data from Heithausen \& Thaddeus 1990 and
Falgarone et al. 1998) and $^{13}$CO (data from Heyer \& Schloerb 1997)
lines and it looks that power law spectra are quite generic for 
interstellar turbulence (Armstrong et al. 1997). Thus the assumption 
of a power law statistics
does not tangibly  constrain the range of applicability of the developed 
theory\footnote{It is rather unnatural to expect that velocity and
density spectra not being power laws conspire to produce power law
emissivity.}.  

For power-law spectra of density with $n>-3$ 
the correlation functions are also power-law:
\begin{equation}
\xi(r)=\langle \rho \rangle^2 
\left(1 + \left( {r_0 \over r} \right)^\gamma\right), ~~~~~ \gamma=n+3 > 0~~~.
\label{eq:xi}
\end{equation}

Substituting Eq.~(\ref{eq:xi}) in (\ref{eq:main}) one can see that 
\be
P_s(|{\bf K}|,k_z)=\langle \rho \rangle^2
\left[P_{v}(|{\bf K}|,k_z)+P_{\rho}(|{\bf K}|,k_z)\right] ~~~, 
\label{eq:split}
\ee
where the part $P_v$ comes from integrating unity in Eq.~(\ref{eq:xi})
and the part $P_{\rho}$ comes from integration the $\left( {r_0 \over r} 
\right)^\gamma$ part. As we may see, the $P_{\rho}$ part is influenced
by both velocity and density fluctuations, while $P_v$ part arises
only from density fluctuations. LP99 show that an expression is 
similar to (\ref{eq:split}) valid for
$n<-3$.

The relation between 2D spectrum in velocity
slices and the underlying 3D emissivity spectrum in the PPV space is given
by  
\be
P_2({\bf K})|_{\cal L} \sim
{1 \over 2 \pi} \int_{-\infty}^{\infty} dk_z  P_s({\bf K}, k_z)\,
2 \left[(1-\cos(k_z {\cal L}) ) / (k_z {\cal L})^2\right] ~~~,
\label{eq:slice}
\ee
where ${\bf K}$ denotes a 2D wavevector defined in the 
planes perpendicular
to the line-of-sight and $|_{\cal L}$ reflects the dependence on the 
slice thickness.
Equation (\ref{eq:slice}) 
represents the one dimensional integral of the three dimensional spectrum
with the window function given by the expression in square brackets.
It is easy to see that the thinner is 
the velocity slice ${\cal L}$, the larger the $k_z$ range for which
the window function is close to unity
and therefore 
more 3D modes contribute to 2D spectrum. 

Substituting (\ref{eq:split}) into (\ref{eq:slice}) one can see that
the two dimensional spectrum can be presented as a sum
\begin{equation}
P_2(|{\bf K}|)=\langle \rho \rangle^2
\left[P_{2v}(|{\bf K}|)+P_{2\rho}(|{\bf K}|)\right]~~~,
\label{2Dspec}
\end{equation}
where the expressions for $P_{2v}$ and $P_{2\rho}$ are self-evident.
To avoid possible misunderstanding 
I would like to stress that  $P_{2v}$ and $P_{2\rho}$
are {\it not} 2D velocity and density spectra and, for instance,
 $P_{2\rho}$ depends both on velocity and density statistics.

Velocity fluctuations are most important for supersonic turbulence
which is the case for cold HI. In this situation the following
power-law asymptotics can be obtained 
(see Table~1):
\be
{\rm thin~~slice}: ~~~~~~~~ C  |{\bf K}|^{-m} \gg \delta V^2 
\label{eq:phystrans1}
\ee
\be
{\rm thick~slice}: ~~~~~~~~ C  |{\bf K}|^{-m} \ll \delta V^2
\label{eq:phystrans}
\ee
In other words, if the velocity dispersion $Cr^{m}$ on the 
scale $|{\bf K}|^{-1}$
is larger than the squared width of the channel  (in velocity units)
the slice is termed {\it thin}. If the opposite is true  the slice is 
termed {\it thick}.

\begin{table}[h]
\begin{displaymath}
\begin{array}{lrr} \hline\hline\\
& \multicolumn{1}{c}{\rm thick~~slice}  &
\multicolumn{1}{c}{\rm thin~~slice}
\\[2mm] \hline \\
P_{2\rho}({\bf K}): &
|{\bf K}|^n;
&|{\bf K}|^{n+m/2} \\[2mm] 
\hline \\[3mm]
P_{2v}({\bf K}): & |{\bf K}|^{-3-m/2}; &
|{\bf K}|^{-3+m/2} \\[3mm] \hline
\end{array}
\end{displaymath}
\caption{Asymptotics of the  components of 2D spectrum in the {\it thin}
and {\it thick} velocity slices, $m=-\nu-3$.} 
\label{tab:2Dspk_asymp}
\end{table}

{\it Thin Slices}\\
It is easy to see from Table~1 that in thin slices the velocity 
mapping makes the spectra more shallow (as $m>0$). This means that
velocity creates a lot of small scale structures. It is also
evident that if the $n<-3$ the $P_{2v}$ contribution dominates.
In the opposite regime $P_{2\rho}$ contribution is important.

{\it Thick Slices}\\
The relative importance of $P_{2v}$ and $P_{2\rho}$ depends on
whether $n>-3-m/2$ or  $n<-3-m/2$. In the former case $P_{2\rho}$
dominates, while for the latter  $P_{2\rho}$ becomes dominant
only when the slice is {\it very thick}, i.e. a substantial portion
of the line is integrated over. Indeed, it is easy to see
that integration over the line-width leaves only density information.
 In the intermediate situation if the density spectrum is steep, i.e.
$n<-3-m/2$
 $P_{2v}$ provides most of the contribution to $P_2$.

For warm HI the thermal velocity dispersion is comparable with
the turbulent one. Thus fluctuations of intensity arise mostly
from density inhomogeneities and the analysis in L95
is applicable. The amplidude of fluctuations arising from the
warm phase of HI is suppressed due thermal velocity smearing
effects. Therefore in the mixture of the warm and cold phase
the contribution of the cold phase to the measured spectrum
is likely to dominate (LP99).

\section{Statistics of Diffuse HI}

\subsection{Analysis of data}

One of the most thorough jobs of obtaining 21~cm statistics was done
by Green (1993). His observations
of the HI emission were accomplished with the Synthesis
Telescope of the Dominion Radio Astrophysical Observatory (DRAO)
towards $l= 140^{\circ}, b=0^{\circ}$ ($03^{h} 03^{m} 23^{s},
+58^{\circ} 06' 20'$, epoch 1950.0) and they revealed a power law spectrum of
2D intensity. This spectrum is proportional to $P_2({\bf K})|_{\cal L}$
and its interpretation depends on whether the slicing is thick or thin.
To answer this question one has to estimate the dispersion of turbulent
velocity at the scales under study and compare it to the velocity
thickness of the slice (see eq.(\ref{eq:phystrans1}) and (\ref{eq:phystrans}).
Assuming that velocity variations at the scale $30$~pc amount
to $10$~km/s and arise from the Kolmogorov turbulence, the structure
functions of velocity are
\be
D_{LL}(r)\approx 100
\left(\frac{r}{30 {\rm pc}}\right)^{2/3}~{\rm km}^2{\rm s}^{-2}~~~,
\ee
The width of the interferometer channels combined to give
a single data point in Green's dataset is $\delta V =5.94~\mathrm{km/s}$.
The slice thickness in parsec is ${\cal L}\approx \delta V f$~pc, 
and varies from
$\approx 600$~pc for the closest slices to  $\approx 2200$~pc for the 
distant ones\footnote{Note, that the cut-off due to thermal velocity (see
section~2.1) in Warm Neutral Medium (see a table of idealized phases
in Draine \& Lazarian (1999) ) is $\sim 6$~km/s. If the WNM constitutes
the dominant fraction of the neutral phase (Dickey 1995) then the
velocity resolution above is optimal and no further decrease in $\delta V$
will result in getting new information. However, if close to Galactic
plane Cold Neutral Media constitutes a substantial portion of mass, the
increase of velocity resolution up to 1~km/s is desirable.}. The wavenumber of transition from {\it thin} to {\it think}
slice given by eq.~(\ref{eq:phystrans}) is equal $0.16\,pc^{-1}$. 

In figure~1 the turbulence scales covered by Green's study are shown.
The smallest 
$|\bf K|$ span from $\sim 1/3$~pc$^{-1}$ for the closest slices
to $1/200$~pc$^{-1}$ for the distant ones.
It is obvious that most of the measurements correspond to the
thin slice regime. 

\begin{figure}[ht]
\centerline{\psfig{file=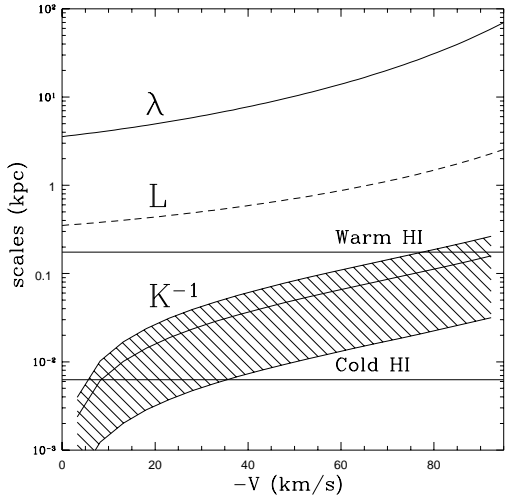,width=7cm,angle=0}}
%\centerline{\epsfxsize=3.in\epsfbox{flat_rot.eps}}
\caption{The variations of geometric scales with the sampling velocity are
shown (from LP99). The upper curve corresponds to the variations of 
the correlation
scale $\lambda=(f^2C)^{1/(2-m)}$ 
in the velocity space. Physically this is the scale at which the
square root of velocity dispersion $\sim (C\lambda^m)^{1/2}$ become
equal to the difference of the regular velocities due to Galactic rotation.
The middle curve corresponds to the variations
of the slice thickness $L$. The darkened area in the Figure depicts
to the range of the turbulence scales under study in Green (1993). 
The solid curve within the darkened area corresponds to the interferometric
measurements with the baseline 21~m. The lower
horizontal line denotes value of
$|{\bf K}|^{-1}$ which separates {\it thin} (above) and {\it thick} slice
regimes (below) for cold HI. The contribution of fluctuations from the
warm phase is suppressed when the slices are thin for cold HI and thick
for warm HI (LP99).}
\label{fig:flat_rot}
\end{figure}
 
As we mentioned earlier, whether $P_{2v}$ or $P_{2\rho}$ dominates
the observed emissivity spectra in the thin slice regime
depends on whether $n$ is larger or smaller than $-3$. If $n>-3$, 
 $P_{2\rho}$ is dominant and the observations by Green (1993) reveal 
the spectrum with index $n+m/2$. For $m=2/3$ the spectrum of emissivity
obtained by Green (1993), namely the emissivity with the index $\sim -2.7$,
corresponds to $n \sim -3$. If, however, the density spectrum is 
steep (i.e. $n<-3$), the
fluctuations of 21~cm intensity observed by Green (1993) can arise
from velocity fluctuations. In this case the spectral index is
$-3+m/2$. For $m=2/3$ one gets the slope $-8/3\approx -2.7$ which
is exactly what is observed. The question now is whether nature
conspires to create the density spectrum with $n\approx -3$ and
thus make the slopes of  $P_{2v}$ or $P_{2\rho}$ identical in
thin slices or we observe  $P_{2\rho}$, while $n<-3$.

To answer this question one has to consider thick slices of data.
Unfortunately, for thick slices of data one has to account for
 lines of sight being not parallel. This problem was studied
by Lazarian (1994) for the case of density statistics, but the
study has not been generalized so far for the case when both density
and velocity contribute to emissivity. Thus external galaxies provide
a better case for study thick slices.  Data for Small
Magellanic Cloud
 (SMC) in Stanimirovic et al. (1999) shows the
steepening of the slope from $\approx -3$ for the slices obtained
with the maximal velocity resolution  to $n\approx-3.5$ for data 
integrated over
the entire emission line
(Stanimirovic, private
communication) which corresponds to our theoretical predictions for
long-wave dominated density spectrum with the index $n \approx -3.5$. 
Thus the set of Green's and Stanimirovic's data is consistent with
an interpretation that both velocity and density exhibit spectra
close to Kolmogorov.

A potential difficulty of ``Kolmogorov'' cascade interpretation is
that the SMC data show power-law slope up to 4~kpc scale. To explain
Doppler broaderning of molecular lines one has to accept that the
energy is being injected at large scales. However scales of
several kpc look excessive.
Injection
of energy at such large scales is possible in the form of superbubles,
but the details and the very possibility of the cascade in these circumstances
is unclear.

\subsection{Further tests}

The ``Kolmogorov'' interpretation above apparently needs further testings.
There are various pieces of evidence that could be
interpreted in favor of the
spectrum of density being shallow, i.e. $ \approx -3$. 
However, our analysis shows that this interpretation is not substantiated.

For instance, Braun (1999) reports a power law
index of $-3$ for the spectrum of 21~cm emission
from structures near the North-East major axis of M31 galaxy. 
However, he uses not the whole spectral lines, as we do in our
treatment but the maximal values of velocity only 
(Braun 1999, private communication). The interpretation
of this result in terms of the power spectra as discussed in L95 and
LP99 is impossible, as the treatment of data is very different.

Shallow spectrum of Far Infrared emission
from dust (Wall \& Waller 1998, Waller et al 1998) does not support the
the shallow HI density spectrum either. According to Stanimirovic (1999,
private communication) the shallow spectrum of Far Infrared emission
when converted into dust column density provides a steep ``Kolmogorov''-type
spectrum. 

One can argue that a 
possible hint in favor of HI density being short wave dominated comes from
molecular data discussed in Stutzki et al. (1998). There  for both
 $^{12}$CO (data from Heithausen \& Thaddeus (1990) and
Falgarone et al. (1998))
and $^{13}$CO (data from Heyer \& Schloerb (1997))
transitions the spectrum of intensity was observed
to have a power law index $\sim -2.8$. As the data is averaged over
velocity, the fluctuations of intensity are due to density fluctuations
and the spectrum of density should have the same
slope as the spectrum of emissivity (L95), provided
that the transitions are optically thin. The problem is that they are
not thin and therefore the analysis above is not applicable.

Attempting to establish the actual underlying spectrum one may 
use one dimensional spectrum introduced in LP99
\begin{equation}
P_1(k_z) = \int d{\bf K} P_s({\bf K}, k_z)
\label{16}
\end{equation}
Similar to two dimensional spectrum $P_1$ can be presented  as a sum of
$P_{1\rho}$, which scales as $k_z^{2(n+2)/m}$ and $P_{1v}$, which
scales as $k_z^{-2/m}$. Naturally, if $n<-3$ $P_{1v}$ dominates,
while $P_{1\rho}$ dominates for $n>-3$. The analysis of data using 
$P_1(k_z)$ has not been done so far.

\section{Anisotropies and Magnetic field}

\subsection{Goldreich-Shridhar Turbulence}

It is natural to expect that dynamically important magnetic field makes
interstellar turbulence anisotropic (Montgomery 1982, Higton 1984). Indeed, 
it gets difficult for
hydrodynamic motions to bend magnetic fields at small scales if
the energy density of the magnetic field and hydrodynamic motions 
are comparable at large scales.
The turbulence in ionized gas
has been found to be anisotropic and its  Kolmogorov-type spectrum of 
plasma density fluctuations
observed via radio scintillations and scattering (see Armstrong 
et al. 1995 and references therein) has been interpreted recently 
as a consequence of
a new type of MHD cascade by Goldreich \& Sridhar (1995).
 The Goldreich-Shridhar model of turbulence\footnote{A qualitative
discussion of the model and the role of reconnection for the
cascade can be found in Lazarian \& Vishniac (1999).} differs considerably
from the Kraichnan one (Iroshnikov 1963, Kraichnan 1964).
It accounts for the fact that hydrodynamic motions can easily mix up
magnetic field lines in the plane perpendicular to the direction of
the mean field (see discussion in Lazarian \& Vishniac 1999). 
Such motions provide eddies elongated in the 
field direction and the velocity spectrum close to the Kolmogorov one.

The Goldreich-Shridhar turbulence is anisotropic with eddies stretched 
along magnetic
field. 
The wavevector component parallel to magnetic field $k_{\|}$
scales as $k_{\bot}^{2/3}$, where $k_{\bot}$ is a wavevector component 
perpendicular
to the field. Thus the degree of anisotropy increases with the
decrease of  scale. 

\subsection{Anisotropies and magnetic field direction}

It is both  challenging and important to determine
the degree of anisotropy for the HI statistics for various parts of
the Galaxy. This information can provide an insight
to the nature of HI turbulence and may be used as a diagnostic for the
interstellar magnetic field. For instance, measurements  of
 the structure functions
of HI intensity 
\be
S(\theta,\phi)=\langle (I({\bf e}_1)-I({\bf e}_2))^2\rangle
\ee 
as a function of a positional angle $\phi$ for individual
subsets of data
should reveal magnetic field direction in various portions of the sky, 
if  the turbulence is anisotropic as we expect it to be. This technique 
is somewhat analogous to a technique of finding magnetic field direction
using the fluctuations of synchrotron radiation (see Lazarian 1992) but
its applicability may be much wider. 

So far, the attempts to measure anisotropy in HI are limited to the 
Green (1994) study,
where no anisotropy was detected. Apparently a better analysis is
needed. For the slices with high degree of
anisotropy the statistical technique can be improved as suggested in L95. 

\section{Structures in PPV space}

PPV data cubes, e.g. HI data cubes, exhibit a lot of small scale 
emissivity structure\footnote{It was noticed by Langer, 
Wilson \& Anderson (1993) that more structure
is seen in PPV space than in the integrated intensity maps.}. The question
is what part of them is real, i.e. is associated with density enhancements
in galactic coordinates and what part of them is produced by velocity 
fluctuations. A related question is whether the structures we see
 are produced dynamically, through
forces, e.g. self-gravity, acting on the media or they may be produced
statistically exhibiting the properties of random field. The second
question was partially answered in Lazarian \& Pogosyan (1997), where it 
was shown that density fluctuations with Gaussian distribution and
power spectra result in filamentary structures. The structures become
anisotropic and directed towards the observer when the velocity effects
are accounted for\footnote{There is a distant analogy between this effect
and the ``fingers of God'' effect (see Peebles 1971) in the
studies of large scale structure of the Universe.} (see Lazarian 1999, fig.~2). 

The issue of density enhancements produced by velocity fluctuations is
closely related to the statistics of ``clouds'' observed in
PPV space. The results on velocity mapping that we discussed earlier
suggest that spectra of fluctuations observed in PPV velocity slices
are more shallow than the underlying spectra. This means
more power on small scales 
or, in other words, more small scale structure (``clouds'')  
appears in the PPV slices due to velocity fluctuations.

It has been believed for decades that emission cloud surveys 
(see Casoli et al. 1984, Sanders et al. 1985, Brand \& Wouterloot 1995)
provide a
better handle on the actual spectrum of cloud mass and sizes than
the extinction surveys (see Scalo 1985) because
the velocity resolution is available. These two sorts of survey 
present different slopes for clump sizes and the difference cannot
be accounted through occlusion of small clouds in the extinction surveys
by larger ones (Scalo \& Lazarian 1996). In view of the discussion above
it looks that extinction survey may be closer to the truth, while a lot of
structure detected via analyzing PPV cubes is due to velocity caustics.
Paradoxically enough, emission data integrated over the spectral lines
may provide a better handle on the distribution of cloud sizes compared
to high resolution spectral-line data cubes. Averaging over velocity
 results in the distortions of the cloud size spectra due to occlusion
effects, but these effects can be 
accounted for using the formulae from Scalo \& Lazarian (1996).        

\section{Generalization of the technique}

The formalism was described above in terms of HI power law statistics.
It is obvious that it can be modified to deal with arbitrary statistics
and with a variety of emission transitions. Here we briefly discuss
complications which a generalization of the technique in order to be
applicable to molecular clouds and ionized media may encounter. 

\subsection{Forward and inverse problems}

A considerable number of researchers believes that self-similar behavior 
reflected in power law statistics is a characteristic feature of the
interstellar turbulence including the molecular cloud turbulence
(see Elmegreen \& Falgarone 1996). However, 
some researchers (e.g. Williams 1999) see departures from a
power law, e.g. signatures of a characteristic scale. In those cases,
one still can find the underlying density and velocity 
statistics solving {\it forward} and {\it inverse} problems (see
Lazarian 1999).

To solve the forward problem one needs to use expressions for the observable
statistics, e.g. expressions for 2D and 1D spectra in PPV space (see
eqs~(\ref{eq:main}) 
and (\ref{16})) and fit the observable statistics varying the input of the
velocity and density statistics. Naturally, the question of uniqueness
for such solutions emerge, but with a reasonable choice of input 
parameters one may hope to avoid ambiguities. 

A different approach involves the inversion of input data. 
Inversion also requires a model, but for the case of turbulence
studies the model can be quite general and usually includes
some symmetry assumptions, like the assumption of isotropy or
axial symmetry of turbulence statistics (Lazarian 1994a, L95). For the
case of turbulence the inversion has been developed for statistics
of density (L95, Lazarian 1993) and magnetic field (Lazarian 1992, 
Lazarian 1994a). A remarkable property of
the inversion for turbulence statistics is that it allows analytical
solutions, which shows that the inversion is a well posed procedure 
in the mathematical sense. The procedure for inverting velocity
data should be analogous to inverting density \& magnetic field
statistics, but has not been developed so far.
We expect  wide application of forward and especially
inverse problem technique
when the deviations from self-similar behavior become apparent in the
data. 

\subsection{Various Transitions}

As we discussed earlier, one of the advantages of using HI as a test
case is that the emissivity is proportional to the column density. 
This is true for some optically thin transitions in molecular clouds,
but fails when absorption is important. My analysis showed that
the absorption is relatively easy to account for if it arises from
dust, but much harder to deal with if it is self-adsorption. In the
former case most of the analysis above is valid provided that the
turbulence scale under study is much smaller than the extinction
length.

Homogeneity of sample is another major concern for studies of
turbulence in molecular clouds (see Houlahan \& Scalo 1990).
Filtering the data (see Miesch \& Bally 1994), application of
wavelets (see Stutzki et al. 1998) or both are required. However, it seems
that as the resolution of data improves the effects of cloud edges
get less important and easier to take care of.

Some emissivities, e.g. those of H$\alpha$ lines are proportional
to the squared density of species. However, it is possible to generalize 
the technique above 
for those transitions and provide a quantitative treatment of
turbulence in ionized
emitting media, e.g. of HII regions (O'Dell 1986, O'Dell \& Castaneda 1987).

\section{Beyond Power Spectra}

The approach that we discussed so far can be characterized as an
interpretation of the emissivity spectra\footnote{
In fact we do not distinguish between spectra and correlation functions.
The two statistics are related via Fourier transform and provide
an equivalent description. In some particular circumstances one or the
other may be more convenient, however.} in terms of the underlying
statistics of velocity and density.

The advantage of this approach is that no numerical 
inversion (see Lazarian 1999) is performed and thus one should not
worry about increase of the data noise. The power spectra are
widely used in hydrodynamics 
and therefore there is hope to relate the  statistics
of interstellar turbulence with something simple and
more familiar like 
Kolmogorov-type cascade.

In spite of all these advantages, the information that this 
approach can supply us with is limited. Indeed, media clustered by
self-gravity and more diffuse media may have the same index of
power spectrum, while being very different. In general, 
statistical measures borrowed from hydrodynamics may not
be adequate while dealing with
interstellar
media. Indeed, we have to address particular questions,
e.g. the question the identification of star-forming regions,
which are beyond the standard hydrodynamic description.
Therefore
attempts to introduce new descriptors are worthy of high praise.
It may happen that in answering specific questions one has to
use particular descriptors. 

\subsection{Spectral Correlation Function}

A new tool termed ``spectral correlation function''  or SCF has been
recently introduced to the field (Goodman 1999, Rosolowsky et al. 1999). 
It compares neighboring spectra with each other. For this
purpose the following measure is proposed:
\be
S(T_1, T_0)_{s,l}\equiv 1-\left(\frac{D(T_1,T_2)}{s^2\int T_1^2(v)dv+
\int T_0^2(v)dv}\right)~~~,
\ee
where the function
\be
D(T_1,T_2)_{s,l}\equiv \left\{\int [s T_1(v+l)-T_0(v)]^2dv\right\}
\ee
and the parameters $s$ and $l$ can be adjusted. One way to choose
them is to minimize $D$ function. In this case $S$ function
gets sensitive to similarities in the shape of two profiles. 
Fixing $l$, $s$ or both
parameters one can get another 3 function that are also 
sensitive to similarities in amplitude, velocity offset
or to both parameters.

The purpose of those functions is to distinguish regions with various
star forming activity and to compare numerical models with observations.
To do this histograms of SCF are compared with
histograms of SCF obtained for the randomized spatial positions. 
This allows to models to be distinguished on the basis of their clustering
properties. First
results reported by Rosolowsky et al. (1999) are very encouraging. 
It was possible to find differences for simulations corresponding to
magnetized and unmagnetized media and for those data sets for which
an earlier analysis by Falgarone et al. (1994) could not find the difference. 
The mathematical development of this new tool is under way (Padoan et al. 2000)
and we expect new exciting results to be obtained in the
near future.

A few comments about spectral correlation functions may be relevant.
First of all, by its definition it is a very flexible tool. In the
analysis of Rosolowsky et al. (1999) the SCF were calculated for the
subcubes over which the original data was divided. In this way SCF
preserves the spatial information and in some sense is similar
to cloud-finding algorithms (see Stutzki \& Gusten 1990, Williams,
de Geus \& Blitz 1994). However, one may fix the angular separation between
the studied spectra and then the technique will be more similar to
the traditional correlation function analysis that is sensitive to
turbulence scale rather than to positional information. I believe that
this avenue should be explored in future along with other more sophisticated
techniques that can be applied to SCF.  
At first glance, it looks counterproductive to get a whole lot of 
various correlations using SCF as the input data. However,
we must find a way of distinguishing regions with various physical
properties and we are still in search for the best descriptors.

At the moment the distinction between various interstellar regions and
the sets of simulated data is made by eye examining the histograms of
SCF. With more information available it seems feasible to use wavelets
that will emphasize some characteristics of the histograms in order
to make the distinction quantitative. Construction these wavelets
will be the way of ``teaching'' SCF to  extract 
features that distinguish various sets of data.   

\subsection{Genus Statistics}

The topology of ISM is an essential characteristic
of the media. Genus analysis has been proved to be a useful tool for
characterizing topology of the Universe (see Gott et al. 1989) and
therefore it is tempting to apply it to the ISM studies. 

The two dimensional genus analysis can be directly related to the
media topology. By 2D maps we mean here maps integrated over the
emission line, i.e. total intensity maps.

A two-dimensional genus is (Melott et al. 1989)
$$ G_2(\nu_t)=({\rm number~of~ isolated~ high~ density~ regions}) - 
({\rm number~ of~ isolated~ low~ density~ regions})$$
where $\nu_t$ denotes the dependence of genus on the threshold density in units
of standard deviations from the mean. It is obvious that if one raises
the density threshold from the mean,
the low density regions coalesce and the genus
becomes more positive. The opposite is true if $\nu_t$ decreases. Thus
for Gaussian fluctuations one expects genus to be antisymmetric about
zero, but the actual distributions should be able to
 reveal ``bubble'' or ``meatball''
topology of various parts of the 
ISM. Algorithms exist for calculating genus for
2D maps, e.g. microwave background maps (Colley et al. 1996) and
therefore the application of genus statistics to interstellar 
maps is straightforward (and long overdue).

The 3D genus statistics (see Gott et al. 1989)
in PPV space is less easy to interpret. As we
discussed earlier, a lot of structures there are due to velocity caustics
and the relation of the structures in galactic coordinates and PPV space
is not obvious. However, it seems interesting to apply genus analysis
to the PPV space in search for another statistical tool to distinguish
various interstellar regions. After all, SCF introduced by Alyssa Goodman
do not have a straightforward relation to the known parameters, 
but are very useful.

\subsection{Bispectrum}

Attempts to use multipoint statistics are a more traditional way to
remove the constraints that the use of two point statistics, e.g.
power spectra entails. Unfortunately, very high quality data is
needed to obtain the multipoint statistics. 

Among multipoint statistics, bispectrum (see Scoccimarro 1997)
seems the most promising. This is partially because it has been successfully
used in the studies of the Large Scale Structure of the Universe.

Bispectrum is a Fourier transform of the three point correlation
function and if the power spectrum $P({\bf k})$ is defined as
\be
\langle \delta\rho({\bf k}_1)\delta 
\rho({\bf k}_2)\rangle = P({\bf k})\delta_D ({\bf k}_1+{\bf k}_2)
\ee
where $\delta_D$ is the Dirac delta function that is zero apart from
the case when ${\bf k}_1+{\bf k}_2=0$, the bispectrum $B_{123}$
is
\be
\langle \delta\rho({\bf k}_1)\delta \rho({\bf k}_2)\delta \rho({\bf k}_3)
\rangle=B_{123}\delta_D ({\bf k}_1+{\bf k}_2+{\bf k}_3)   
\ee

It is advantageous to use 
``hierarchical amplitude'' (Fry \& Seldner 1982) statistics
\be
\Phi_{123}\equiv \frac{B_{123}}{P({\bf k}_1)P({\bf k}_2)+ P({\bf k}_2)
P({\bf k}_3)+P({\bf k}_3) P({\bf k}_1)}
\ee
which is for power law spectra is a scale independent quantity.

In the studies of Large Scale Structure the hierarchical amplitudes
were calculated for various initial conditions to compare with observations.
In interstellar media it is advisable to compare various regions of
sky using the tool. Impediments for the use of the technique stem from
the increase of noise with the use of multipoint statistics and the
problems of averaging along the line of sight. 
The problems should be addressed in the future.

\subsection{Other techniques}

A wavelet technique (see Gill \& Henriksen 1990, Langer, Wilson \&
Anderson 1993, Rauzy, Lachieze-Rey \& Henkiksen 1993) 
is discussed in this volume by Stutzki who
proposed a so called $\Delta$- variance technique (Stutzki et al. 1998)
which is is related to wavelet transforms (Zielinsky \& Stutzki 1999).
Wavelets potentially are a versatile tool that can filter out the
large scale inhomogeneities of the data and concentrate the analysis
on the scales of interest (see Stutzki, this volume).

Another useful statistical tool is the Principal Component Analysis
(PCA). This tool was employed to spectral line imaging studies  of 
the interstellar medium by Heyer \& Schloerb (1997). The goal of the
PCA is to determine a set of orthogonal ``axes'' $u_{kl}$ for which the 
the variance of the data is maximized. In the case of the data in
$n$ points with $p$ velocity (spectrometer) channels for each point
the data can be presented as $\delta T_{ij}=T_{ij}-\langle T_{ij} \rangle_n$, 
where $T_{ij}$ is the temperature at the channel $j$ at a position $i$
and $\langle ... \rangle_n$ denote averaging over positions. Maximazing
the variance means maximizing the expression $y_{ij} y_{ij}=u_{ik}S_{jk}
u_{ij}$, where summation over the repeating indexes is implied and
$S_{ik}=\langle \delta T_{ij} \delta T_{jk} \rangle_n$. In practice finding of
$u_{ij}$ amounts to solving a set of eigenvalue equations 
$S_{ik}u_{kj}=\lambda u_{ij}$. To visualize the variance related to $l$-th 
principal component eigenimages are constructed from the projections of 
$T_{ij}$ onto the eigenvector, i.e. $l$th eigenimage at point
 $(r_i)$ is $\delta T_{ij}u_{l,j}$.  
Heyer \& Schloerb (1997) showed that using PCA technique it is possible
to decompose large-scale spectroscopic images of molecular clouds. Their
analysis enabled them to calculate the velocity - scale relations for
a number of cloud complexes. In terms of the statistical analysis
presented above, PCA provides a means of filtering out large scale
features responsible for the largest contribution to the global variance.
This makes the sample more homogeneous and suitable for describing using
correlation functions and power spectra. The potential of this 
statistical tool is to be further explored. It is likely that combining
the various set of data (for instance, HI and CO) more interesting
correlations can be obtained via PCA technique.

\section{Discussion}

It is not possible to cover all the various interesting approaches
that have been tried in order to study interstellar turbulence via emission
lines. For instance, we omitted a discussion of
 3D correlation functions in PPV space  introduced
in Perault et al. (1986). We did not cover studies of turbulence
using centroids of velocity (see Dickman 1985) either.
One reason for this is that I believe that the statistics of
velocity centroids have to be described in terms of underlying velocity
and density.

A search for tools to deal with the interstellar turbulence has been
intensified recently and this shows 
that there is deep understanding
in the community that the wealth of observational data must be explored
and it is essential to compare observations and numerical simulations.
I personally believe that the development of theoretical approaches
to dealing with data has become at this point not less important than
obtaining the data.

Most of the present review I devoted to dealing with power spectra which
reflects my personal preferences. Although far from being unambiguous,
the power spectra were most intensively studied in hydrodynamics and
the MHD theory and therefore they provide a bridge between  idealizations
that we partially
understand and {\it terra incognita} of  interstellar turbulence.
Whether this approach is useful for a particular phase of the 
interstellar media is not clear {\it a priori}. We may or may not have
any self-similarity indicating a turbulent cascade. However, at least
for HI it seems that the approach is promising. Indeed, we managed
to relate, although tentatively, the statistics of 21~cm emission with
the statistics of Goldreich-Shridhar cascades. The next class of objects
 to study using the technique should be molecular clouds.

Although studies of molecular clouds are expected to face more problems,
some of them mentioned earlier on, it is likely that the underlying 3D
statistics will be soon obtained for the optically thin molecular lines.
Comparison of this statistics with that in diffuse media should provide
an insight to the nature of interstellar turbulent cascade and 
turbulent support of molecular clouds.

However, the limitations of the power-spectrum approach make it necessary
to use alternative tools such as wavelets, genus statistics, principal
component analysis and develop new ones such as spectral correlation function
even though the relation between their output and the familiar
notions from hydrodynamics is not always clear. In studies of
interstellar medium one has  to address particular questions, e.g.
the question of star formation and therefore appearance of specialized
tools is only natural.

\section*{Summary}

~~~~1. Velocity and density power spectra can be obtained
from observed emissivities. Velocity fluctuations make emissivity 
spectra in velocity slices shallower.
This results in much of small scale structure in PPV space
that can be erroneously interpreted as interstellar clouds or clumps. 

2. Turbulence is likely to be anisotropic with magnetic field defining
the anisotropy direction. This should allow a new way of studying 
magnetic field.

3. The available wealth of observational data motivates the development
of new tools for data handling.

\section*{Acknowledgments}

This review is partially based on the results that I obtained together
with Dmitry Pogosyan. Discussions with Chris McKee, 
John Scalo, Steve Shore and Enrique Vazquez-Semadeni
are acknowledged. I am grateful to Robert Braun and Snezana Stanimirovic
for helpful input on data reduction procedures. 

%\section*{Appendix}


\section*{References}


\begin{thebibliography}{99}
\bibitem{} Adler, D. S. \& Roberts, W. W. Jr. \Journal{1992}{\ApJ}{384}{95}
\bibitem{} Armstrong, J.M., Rickett, B.J., \& Spangler, S.R.
\Journal{1995}{\ApJ}{443}{209}
\bibitem{} Brand, P. \& Wouterloot, J.G.A. \Journal{1995}{\AaAp}{303}{851}
\bibitem{} Braun, R. \Journal{1997}{\ApJ}{484}{637}
\bibitem{} Braun, R. (1999) in {\em Interstellar Turbulence}, eds. 
Jose Franco \& 
Alberto Carraminana, CUP, p. 12
\bibitem{}  Burton, W.B. \Journal{1970}{\em A\&AS}{10}{76}
\bibitem{}  Burton, W.B. \Journal{1971}{\AaAp}{10}{76}
\bibitem{}  Burton, W.B. (1992) in  {\em Galactic Interstellar
Medium}, eds. Pfenninger~D., Bartholdi~P., Springer-Verlag, p.~1 
\bibitem{} Casoli, F., Combes, F., \& Gerin, M. \Journal{1984}{\AaAp}{133}{99}
\bibitem{} Colley, W.N., Gott, J.R. \& Park, C. \Journal{1996}{\MNRAS}
{281}{L82} 
\bibitem{} Dickey, J.M.,  (1995),  in {\em The Physics of the 
Interstellar Medium
and Intergalactic Medium},  eds Ferrara~A., McKee~C.F., Heiles~C.,
Schapiro~P.R., ASP Conf. Ser. V80.
\bibitem{} Dickman, R.L.,  (1985),  
in  {\em Protostars and Planets II}, eds Black~D.C. and Mathews~M.S., 
Tucson:  University of Arizona, 150
\bibitem{}   Dickman, R.L. \& Kleiner, S.C. \Journal{1985}{\ApJ}{295}{479}
\bibitem{} Draine B.T. \& Lazarian, A. (1999), in {\em Sloan Summit 
on Microwave 
Foregrounds}, ed. A. de Oliveira-Costa and M. Tegmark, in press
\bibitem{} Elmegreen, B.G. \& Falgarone, E. \Journal{1996}{\ApJ}{471}{816}
\bibitem{} Falgarone, E., J.-L. Puget, \& Perault, M. \Journal{1992}{\AaAp}
{257}{715}
\bibitem{} Falgarone, E. et al.  \Journal{1998}{\AaAp}{331}{669}
\bibitem{} Falgarone, E., Lis, D. C.,
 Phillips, T. G., Pouquet, A.;
 Porter, D. H., Woodward, P. R. \Journal{1994}{\ApJ}{436}{728}
\bibitem{}  Gautier, T.N., Boulanger, 
F., P\'{e}rault, M. \& Puget, J.-L. \Journal{1992}{\AJ}{103}{1313}
\bibitem{} Gill, A.G. \& Henriksen, R.N. \Journal{1990}{\ApJ}{365}{L27}
\bibitem{} Goldreich, P. \& Sridhar, S. \Journal{1995}{\ApJ}{438}{763}
\bibitem{} Goodman, A.A. 1999, 
http://cfa-www.harvard.edu/~agoodman/scf/SCF/scfmain.html
\bibitem{} Goodman, J., \& Narayan, R. \Journal{1985}{\MNRAS}{214}{519}
\bibitem{} Gott, J.R., Mao, S., \& Park, C. \Journal{1992}{\ApJ}{385}{26}
\bibitem{} Gott, J.R., Melott, A.L., \& Dickinson, M. \Journal{1986}{\ApJ}{306}
{341}
\bibitem{} Gott, J.R., Park, C., Juskiewicz, R., Bies, W.E., Bennett,
D.P., Bouchet, F.R., Stebbins, A. \Journal{1990}{\ApJ}{352}{1}
\bibitem{} Gredel, R. (1999) in 
{\em Interstellar Turbulence}, eds. Jose Franco \& Alberto Carraminana, CUP,
140
\bibitem{} Green, D.A. \Journal{1993}{\MNRAS}{262}{328}
\bibitem{} Green, D.A. \Journal{1994}{\em Ap\&SS}{216}{61}
\bibitem{} Hamilton, A.J.S. 1998, in {\em The Evolving Eniverse}, ed.
D. Hamilton, Dordrecht: Kluwer, p. 185
\bibitem{} Heithausen, A., \& Thaddeus, P. \Journal{1990}{\ApJ}{353}{L49}
\bibitem{} Henriksen, R.N. \Journal{1994}{\em Ap\&SS}{221}{25}
\bibitem{} Heyer, M.H., \& Schloerb, F.P. \Journal{1997}{\ApJ}{475}{173}
\bibitem{} Higgs, L.A. (1999), in {\em New Perspectives on
the Interstellar Medium}, ASP 168, p. 15
\bibitem{} Higdon, J.C. \Journal{1984}{\ApJ}{285}{109}
\bibitem{} von Horner, S. \Journal{1951}{\em Zs.F. Ap.}{30}{17}
\bibitem{} Houlahan, P., Scalo J., \Journal{1990}{\ApJ}{72}{133}
\bibitem{} Iroshnikov, P.S. \Journal{1963}, {\em AZh} {40} {742}
\bibitem{} Kaiser, N. \Journal{1987}{\MNRAS}{227}{1}
\bibitem{} Kalberla, P.M.W. \& Mebold, U. \Journal{1983} 
{\em Mitt Astr. Ges.}{58}{101}
\bibitem{} Kalberla, P.M.W. \& Stenholm, L.G. \Journal{1983}
{\it Mitt. Astron. Ges.} {60} {397}
\bibitem{} Kamp\'{e} de F\'{e}riet, J. (1955), 
in: {\em Gas Dynamics of Cosmic Clouds}, Amsterdam: North-Holland, 134
\bibitem{}  Kaplan, S.A. \& Pickelner, S.B. (1970), 
 {\em The Interstellar Medium}, (Harvard University Press)
\bibitem{}  Kerr, F.J. \& Lynden-Bell, D. \Journal{1986}{\MNRAS}{221}{1023}
\bibitem{} Kolmogorov, A.  \Journal{1941}{\em Compt. Rend. Acad. Sci. USSR}{30}{301}  
\bibitem{}  Kraichnan, R.H. \Journal{1965}{\em Phys. Fluids}{8}{1385}
\bibitem{} Kramer, C., Stutzki, J., Rohrig, R., \& Corneliussen, U.
1998{\AaAp}{329}{249}
\bibitem{} Langer, W.D., Wilson, R.W., \& Anderson, C.H. 
\Journal{1993}{\ApJ}{408}{L45}
\bibitem{} Langer, W. D., Velusamy, T., Kuiper, T.B.H.,
 Levin, S., \& Olsen, E. \Journal{1995}{\ApJ}{453}{293}
\bibitem{} Larson, R. \Journal{1992}{\MNRAS}{26}{641} 
\bibitem{} Lazarian, A. \Journal{1992}{\em Astron. and Astrophys. 
Transactions}{3}{33}
\bibitem{} Lazarian, A. \Journal{1994a}{\em Plasma Phys. and Contr. Fusion}
{36}{1013}
\bibitem{} Lazarian, A. (1994b), {\em PhD Thesis}, University of Cambridge
\bibitem{} Lazarian, A. \Journal{1995}{\AaAp}{293}{507} (L95)
\bibitem{} Lazarian, A. (1999), in 
{\em Interstellar Turbulence}, eds. Jose Franco \& Alberto Carraminana, CUP, 
p.95
\bibitem{} Lazarian, A. \& Pogosyan, D. \Journal{1997}{\ApJ}{491}{200}
\bibitem{} Lazarian, A. \& Pogosyan, D. \Journal{1999}{\ApJ}{}{in press (LP99),
astro-ph/9901241}
\bibitem{} Lazarian, A., \& Vishniac, E. \Journal{1999}{\ApJ}{517}{700}
\bibitem{} Liszt, H.S., \& Burton, W.B. \Journal{1981}{\ApJ}{243}{778}
\bibitem{} McKee, C.F. \&  Ostriker, J.P. \Journal{1977}{\ApJ} {218} {148}
\bibitem{} Melott, A.L., Cohen, A.P., Hamilton, A.J.S., Gott, J.R., Weinberg,
D.H. \Journal{1989}{\ApJ}{345}{618}
\bibitem{} Miesch, M.S., \& Bally, J. \Journal{1994}{\ApJ}{429}{645}
\bibitem{} Montgomery, D. \Journal{1982}{Phys. Scripta}{2}{83}
\bibitem{} Monin, A.S., \& Yaglom, A.M. (1975), 
 {\em Statistical Fluid Mechanics: Mechanics of Turbulence}, vol. 2, The MIT Press 
\bibitem{} Munch, G. \Journal{1958}{\em Rev. Mod. Phys.}{30}{1035}
\bibitem{} Narayan, R. \Journal{1992}{\em Phil. Trans. Royal Soc.}{341}{151}
\bibitem{} Narayan, R., \& Goodman, J. \Journal{1989}{\MNRAS}{238}{963}
\bibitem{} O'Dell, C.R. \Journal{1986}{\ApJ}{304}{767}
\bibitem{} O'Dell, C.R., \& Castaneda, H.O. \Journal{1987}{\ApJ}{317}{686}
\bibitem{} Ostriker, E. (1999) in 
{\em Interstellar Turbulence}, eds. Jose Franco \& Alberto Carraminana, CUP, 
p.~240
\bibitem{} Oughton,S., Radler, K.H., Matthaeus,W.H. \Journal{1997}
{\em Phys. Rev. E}{56}{No.3}, 2875
\bibitem{} Padoan, P., Rosolowsky, E.W., \& Goodman, A.A. \Journal{1999}{\ApJ}
{}{submitted}
\bibitem{} Perault, M., Falgarone, E. \& Puget, J.L. \Journal{ 1986}{\AaAp}
{157}{139}
\bibitem{} Pichardo, B., Vazquez-Semadeni, E., Gazol, A., Passot, T.,
Ballesteros-Paredes, J. \Journal{1999}{\ApJ}{}{submitted}
\bibitem{} Prunet, S. \& Lazarian, A. (1999), in 
``Sloan Summit on Microwave Foregrounds'', ed. A. de Oliveira-Costa 
and M. Tegmark
\bibitem{} Rauzy, S., Lachieze-Rey, M.,\& Henriksen, R.N. \Journal{1993}
{\AaAp}{273}{366}
\bibitem{} Rosolowsky, E.W., Goodman, A.A., Wilner, D.J., \& Williams, J.P.
\Journal{1999}{\ApJ}{}{in press}
\bibitem{} Sanders, D.B., Scoville, N.Z., \& Solomon, P.M. \Journal{1985}
{\ApJ}{289}{373}
\bibitem{} Scalo, J.M. (1985), in {\em Protostars and Planets II}, 
eds Black~D.C. and Mathews~M.S.,  Tucson: University of Arizona, 201
\bibitem{} Scalo, J.M. (1987), in {\em Interstellar Processes}, 
eds. D.F. Hollenbach \& H.A. Thronson, Reidel, Dordreicht, 349
\bibitem{} Scalo, J.M., Lazarian, A. \Journal{1996}{\ApJ}{469}{189}
\bibitem{} Scoccimarro, R. \Journal{1997}{\ApJ}{487}{1}
\bibitem{} Sethi, S.K., Prunet, S., Bouchet, F.R., 1998, astro-ph/9803158.
\bibitem{} Shull, J.M. 1987, in {\it Interstellar Processes}, 
eds Hollenbach~D.J. and Thronson~H.A., 
 Reidel, Dordrecht, 225
\bibitem{} Spangler, S.R., \& Gwinn, C.R. \Journal{1990}{\ApJ}{353}{L29}
\bibitem{} Stanimirovic, S., Staveley-Smith, L., Dickey, J.M.,
Sault, R.J., \& Snowden, S.L. \Journal{1999}{\MNRAS}{302}{417}
\bibitem{} Stenholm, L.G. \Journal{1999}{\AaAp}{232}{495}  
\bibitem{} Stutzki, J., Bensch, F., Heithausen, A., Ossenkopf, V.,
\& Zielinsky, M. \Journal{1998}{\AaAp}{336}{697}
\bibitem{} Stutzki, J., \& Gusten, R. \Journal{1990}{\ApJ}{356}{513}
\bibitem{} Verschuur, G.L. \Journal{1991a}{\em Ap \& SS}{185}{137}
\bibitem{} Verschuur, G.L. \Journal{1991b}{\em Ap \&SS}{185}{305}
\bibitem{} Verschuur, G.L. \Journal{1995}{\em Ap\&SS}{227}{187}
\bibitem{} Vazquez-Semadeni, E. (1999), in  the Proceedings of 
"The Chaotic Universe", Roma/Pescara, Italy, 1-5 Feb. 1999, eds. 
V. Gurzadyan and L.
    Bertone
\bibitem{} Vazquez-Semadeni, E., \& Passot, T. 
(1999),
in {\em Interstellar Turbulence}, eds. Jose Franco \& 
Alberto Carraminana, CUP, p.~223
\bibitem{} Wall, W.F., \& Waller, W.H. 1998, in {\em New Horisons 
from Multi-Wavelength Sky Surveys}, eds. B.J. McLean, D.A. Golombek,
Hayes, J.J.E. and H.E. Payne, IAU 179,
Kluwer, p. 191
\bibitem{} Waller, W.H., Varoxi, F., Boulanger, F., \& Digel, S.W.
 1998,  in {\em New Horisons 
from Multi-Wavelength Sky Surveys}, eds. B.J. McLean, D.A. Golombek,
Hayes, J.J.E. and H.E. Payne, IAU 179,
Kluwer, p. 194 
\bibitem{} Williams, J. (1999) in 
{\em Interstellar Turbulence}, eds. Jose Franco \& 
Alberto Carraminana, CUP, p.~190
\bibitem{} Williams, J., de Gaus, E., \& Blitz, L. 
\Journal{1994}{\ApJ}{428}{693}
\bibitem{} Wilson, O.C., Munch, G., Flather, E.M., \& Coffeen, M.F. 
\Journal{1959} {\em ApJS}{4}{199}
\bibitem{} Wiserman, J.J. \& Ho, P.T.P. \Journal{1996}{\Nat}{382}{139}
\bibitem{} Zielinsky, M. \& Stutzki, J. \Journal{1999}{\AaAp}{347}{630}
\end{thebibliography}
\end{document}